\documentclass[conference]{IEEEtran}
\IEEEoverridecommandlockouts
\usepackage{cite}
\usepackage{amsmath,amssymb,amsfonts}
\usepackage{algorithmic}
\usepackage{graphicx}
\usepackage{textcomp}
\usepackage{xcolor}
\usepackage{url}
\usepackage{verbatim}
\usepackage{xcolor}
\usepackage{multirow}
\usepackage{siunitx}

\usepackage{alltt}
\usepackage{tcolorbox}
\usepackage{balance}
\tcbuselibrary{listingsutf8}
\usepackage{verbatim} % for verbatim environment
\def\BibTeX{{\rm B\kern-.05em{\sc i\kern-.025em b}\kern-.08em
    T\kern-.1667em\lower.7ex\hbox{E}\kern-.125emX}}
\begin{document}

%\title{Passenger-side View of Starlink In-Flight Internet Performance:\\ }
%\title{A First Empirical Look at Starlink In-Flight Internet Performance}
\title{A First Look at Starlink In-Flight~Performance:\\ An Intercontinental Empirical Study}

\author{\IEEEauthorblockN{
Muhammad Asad Ullah\IEEEauthorrefmark{1},
Luca Borgianni\IEEEauthorrefmark{2}, Heikki Kokkinen\IEEEauthorrefmark{1}, Antti Anttonen\IEEEauthorrefmark{1}, Stefano Giordano\IEEEauthorrefmark{2}\\
}
\IEEEauthorblockA{\IEEEauthorrefmark{1}VTT Technical Research Centre of Finland Ltd., Finland}
\IEEEauthorblockA{\IEEEauthorrefmark{2}University of Pisa, Italy}}

%\markboth{IEEE NETWORKING LETTERS,~Vol.~X, No.~X, August~2025}%


\maketitle

\begin{abstract}
%Starlink delivers Internet connectivity to users across terrestrial, maritime, and aeronautical domains. Earlier studies have extensively measured and examined the network characteristics for fixed sites and vehicles in motion. Recently, major airlines partnered with Starlink to offer in-flight high-speed Internet, enabling passengers to browse, stream, and access real-time applications seamlessly. However, there remains a need for scientific research to investigate the Starlink in-flight connectivity performance. 

Starlink delivers Internet services to users across terrestrial, maritime, and aviation domains. The prior works have studied its performance at fixed sites and in-motion vehicles, while an in-depth analysis of in-flight performance remains absent.  With major airlines now offering Starlink Internet onboard, there is a growing need to evaluate and improve its performance for aviation users. This paper addresses this shortcoming by conducting in-flight measurements over the Baltic Sea and the Pacific Ocean. Our measurement results show that a single user device experiences median throughputs of 64~$\mathrm{Mbps}$ and 24~$\mathrm{Mbps}$ for the downlink and uplink, respectively. The median uplink throughput is approximately 33~$\mathrm{Mbps}$ when the aircraft maintains an altitude above 17,000~$\mathrm{feet}$. However, a significant reduction in uplink performance is observed during the aircraft descent phase, with the median throughput dropping to around 20~$\mathrm{Mbps}$ at lower altitudes. Round-trip time (RTT) is highly dependent on the location of the ground station being pinged and the use of inter-satellite links (ISLs). We dive deeper into 5.5 hours of ping measurements collected over the Pacific Ocean and investigate factors influencing RTT, hypothesizing that ISLs routing, data queuing at satellites, and feeder link congestion contribute to deviations from theoretical values. For comparative analysis, we evaluate the Starlink ground terminal and in-flight connectivity performance from the perspectives of a residential user and an airline passenger, respectively.
\end{abstract}
%a median downlink throughput of 64~$\mathrm{Mbps}$  and a median uplink of 24 ~$\mathrm{Mbps}$. 

\section{Introduction}
\IEEEPARstart{I}{n}  the new space era, Starlink operates the world's largest low Earth orbit (LEO) constellation, with more than 8,000 satellites as of August 2025. Starlink provides Internet to stationary and in-motion users, including vehicles, vessels, and aircraft. Due to its operation in LEO, the Starlink radio channel is highly dynamic, with its characteristics varying rapidly over time, which leads to fluctuations in communication performance. In the scientific community, multiple measurement campaigns have examined the Starlink performance. For example, the study in \cite{10.1145/3614204.3616108} explores the Starlink variations in terms of throughput, revealing periodic performance degradation occurring roughly every 15 seconds, likely due to satellite handovers. In a broader context, \cite{10.1145/3589334.3645328} provides a global throughput and round-trip time (RTT) analysis by aggregating more than 19 million measurements from 34 countries.%, complemented by controlled experiments. %In \cite{10521263}, Starlink is further explored for its potential in positioning and navigation applications.

An example of the in-motion scenario is presented in \cite{Asad_Starlink,Dominic_2025}, which investigates the impact of vehicle mobility. These measurements show that Starlink can provide high-speed Internet connectivity to moving vehicles. However, performance fluctuates over time due to highly dynamic channel conditions. Particularly, roadside blockages, e.g., trees and bridges, disrupt the connection. Starlink uses Ku-band for communications~\cite{Humphreys}, which is susceptible to rain and cloud attenuation.  Even light rain can decrease the throughput, while moderate rain causes momentary connection outage~\cite{Asad_Starlink_Weather}. Additionally, \cite{mafakheri2023edge} introduces a system design and prototype that combines 5G, multi-access edge computing, and Starlink-based LEO satellite backhaul to support aircraft on-board connectivity for passengers. Their testbed illustrate how a LEO constellation can serve as a low-latency backhaul to terrestrial 5G core networks, enabling high-speed Internet for onboard systems and passengers.

%Starlink uses Ku-band, specifically 14.0-14.5~$\mathrm{GHz}$ for uplink and 10.7-12.7~$\mathrm{GHz}$ for downlink communications 

%While existing literature offers valuable insights into Starlink’s capabilities, no prior study has systematically assessed its in-flight performance.
%intercontinental 
%Today, Starlink aero terminals are already installed on Hawaiian Airlines A330 and A321neo aircraft. The airline is in the process of equipping its Boeing 787 aircraft with aero terminals.
% delivers Internet to the aviation sector, enabling onboard passengers to browse the web, stream content, and use real-time applications \cite{Parada}.

%  It has already been certified for use on major aircraft, and the range of supported aircraft continues to expand. 
The Starlink aero terminal is designed for aviation users. In late 2024, the first flight successfully demonstrated the feasibility of Starlink in-flight connectivity, keeping passengers connected to the Internet. In North America, Hawaiian Airlines was one of the first leading commercial airlines to provide complimentary in-flight Starlink Internet  \cite{ookla2025starlink}.   In early 2025, airBaltic became the first European airline to provide free-of-charge gate-to-gate Starlink in-flight connectivity to its passengers on Airbus A220-300 fleet. In the Middle East and North Africa region,  Qatar Airways is the first major airline to offer complimentary Starlink connectivity onboard \cite{qatar2025starlink}.   Despite these rapid advancements, to our best knowledge, no prior study has measured the network characteristics of Starlink’s in-flight connectivity from a passenger perspective.

This paper addresses this gap with intracontinental and intercontinental in-flight measurements over the Baltic Sea  and Pacific Ocean and answers the following timely questions:

\begin{itemize}
  \item Can Starlink provide high-speed, low-latency Internet to an aircraft cruising at 37,000~$\mathrm{feet}$ and 800 ~$\mathrm{km/h}$?
  \item How stable is Starlink’s in-flight Internet service in terms of throughput and latency over time?
  \item How does aircraft altitude and speed impact in-flight connectivity performance?
  \item Can Starlink support 4K streaming during flight?
\end{itemize}

Moreover, to have a comparison with a static Starlink connection, we collected some measurements in Pisa, Italy. To motivate further research, we have made our measurement dataset
publicly accessible on GitHub [To be added later].

The rest of this paper is organized as follows. Section~\ref{sec:sec2} presents the technical background and Ookla report for in-flight communication performance. In Section~\ref{sec:sec3} and Section~\ref{sec:sec4}, we discuss the experimental setup and the measured dataset, respectively. Finally, we conclude this paper with final remarks and potential research directions in Section~\ref{sec:sec5}.
%Qatar Airways has completed the installation of aero terminals on its Boeing 777 fleet in July 2025 and is currently proceeding with installations on its Airbus A350 aircraft.
%Section~\ref{sec:sec5} discusses the measurement results and the key takeaways. 
\section{Technical background: In-Flight Connectivity}
\label{sec:sec2} 
The previous measurements confirmed that Starlink terrestrial users typically experience downlink throughput between 25~$\mathrm{Mbps}$ and 220$~\mathrm{Mbps}$. Upload throughput falls between 5$~\mathrm{Mbps}$ and 20$~\mathrm{Mbps}$ \cite{Asad_Starlink}. In satellite communications, physical speed-of-light propagation is one of the biggest factors that affect the RTT. A LEO satellite at an altitude of 550$~\mathrm{km}$ has a theoretical RTT, also known as two-way propagation delay between terminal and satellite in the range of 3.6$~\mathrm{ms}$ to 13$~\mathrm{ms}$, respectively.  Additionally, traffic scheduling, queueing, intersatellite links (ISLs), and ground station and points of presence~(PoP) location contribute to the total RTT \cite{10.1145/3589334.3645328}. Bearing this in mind, Starlink RTT for residential users generally ranges from 25 to 60~$\mathrm{ms}$. The users in the United States (U.S.) are experiencing a median peak-hour RTT of 25.7$~\mathrm{ms}$ and median peak-hour downlink of 200$~\mathrm{Mbps}$ as of June 2025\footnote{Network update (https://www.starlink.com/updates/network-update)}.

\begin{table}[!t]
\caption{Major satellite operators and their partner airlines \cite{ookla2025starlink}.\label{tab:tabI}}
\centering
  \resizebox{\columnwidth}{!}{\begin{tabular}{|l|c|}
\hline
\textbf{Satellite } & \textbf{Partner}\\
\textbf{operator} & \textbf{airlines}\\
\hline
%Deutsche Telekom & Air France, Cathay Pacific, Condor, Lufthansa \\
%\hline
Hughes &  Spirit Airlines\\
%\hline
%Inmarsat &Air New Zealand, Qatar Airways\\
\hline
\multirow{2}{*}{Intelsat}&Air Canada, Alaska Airlines, \\&American Airlines, United Airlines\\
%\hline
%MTN Satellite&\multirow{2}{*}{Southwest Airlines}\\
%Communications&\\
%\hline
%Nelco (PAC/Intelsat)&Air India\\
\hline
%\multirow{8}{*}{\shortstack{Panasonic\\Avionics\\Corporation}}&Aer Lingus, Air France, American Airlines, ANA, \\
%& Asiana Airlines, British Airways, Etihad Airways, \\&EVA Air, Fiji Airways, Finnair, Iberia Airlines, \\& ITA Airways, Japan Airlines, KLM, Korean Air, \\&  Malaysian Airlines,  Scandinavian Airlines,\\& Singapore Airlines, SWISS Airlines,\\& TAP Air Portugal, Thai Airlines, United Airlines, \\&Virgin Atlantic, VoeAzul, WestJet, Zipair Tokyo\\
%\hline
%SITA Switzerland&Qatar Airways\\
%\hline
Starlink&airBaltic, Hawaiian Airlines, Qatar Airways\\
%\hline
%Türk Telekom&	Turkish Airlines\\
\hline
\multirow{4}{*}{Viasat}&Aeromexico, American Airlines, Breeze Airlines,\\& Delta Airlines, EL AL Airlines, Icelandair, \\&JetBlue, Southwest Airlines, United Airlines, \\&Virgin Atlantic\\
\hline
\end{tabular}}
\label{tab:tabI}
\end{table}

Similar to Starlink terrestrial terminals, the aero terminal uses an electronically-steered phased array antenna. The downlink communication channels use 10.7-12.7~$\mathrm{GHz}$, and the uplink channels use 14-14.5~$\mathrm{GHz}$ \cite{Humphreys}. A single downlink and uplink channel has a bandwidth of 250~$\mathrm{MHz}$ and 62.5~$\mathrm{MHz}$, respectively \cite{Asad_Starlink_Weather}. Each Starlink aero terminal has downlink throughput between 100 $\mathrm{Mbps}$ and 250 $\mathrm{Mbps}$, uplink in the range of 8 $\mathrm{Mbps}$ to 25 $\mathrm{Mbps}$, and RTT less than 99~$\mathrm{ms}$. Qatar Airways advertises a throughput of up to 500 $\mathrm{Mbps}$ per aircraft. We assume that each aircraft may be equipped with two aero terminals. In contrast, airBaltic is advertising throughput of up to 350 $\mathrm{Mbps}$, which is consistent with the throughput reported in~\cite{Parada}.  %As a service plan, a 20 $\mathrm{GB}$ aviation data subscription costs \$2,000 per month, where an unlimited data subscription costs \$10,000 per month\footnote{Starlink for Aviation (https://www.starlink.com/fi/business/aviation).}. 

Table~\ref{tab:tabI} lists the satellite network operators and their associated airlines. Starlink is being used by major airlines, including airBaltic, Hawaiian Airlines, and Qatar Airways. Other major air carriers, including United Airlines, SAS (Scandinavian Airlines), Air France, WestJet, and Air New Zealand, have already confirmed plans to adopt Starlink for in-flight connectivity. Unlike Starlink, Hughes collaborates with SES to provide Internet services using medium Earth orbit~(MEO) and geostationary orbit~(GEO) satellites. Similar to Hughes, Intelsat follows a multi-orbit solution that combines LEO and GEO satellites. Viasat uses the GEO constellation for the Internet. In 2025, Viasat has the highest number of partner airlines compared to any other satellite network operator.
%Viasat\footnote{In 2023, Viasat acquired Inmarsat.}

\begin{figure}[!t]
\centering
\includegraphics*[width=0.5\textwidth]{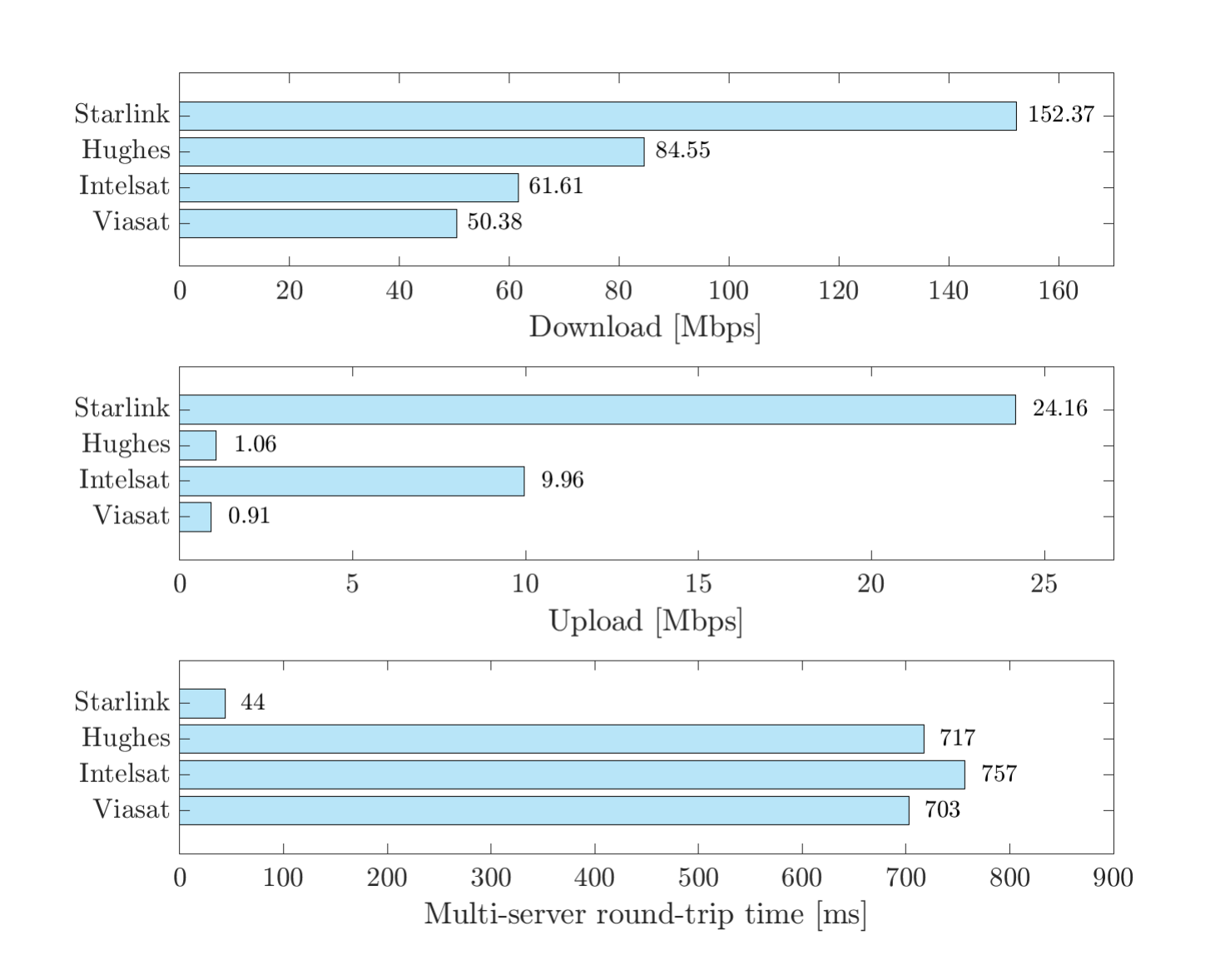}
\caption{Satellite operators and their in-flight connectivity performance.}
\label{fig_1}
\end{figure}

Fig. \ref{fig_1} illustrates different satellite Internet providers and their in-flight connectivity performance. The throughput and RTT data are adopted from the Ookla Speedtest Q1 2025 report \cite{ookla2025starlink}. Starlink stands out in terms of high throughput and low RTT. Particularly, Starlink's median downlink and uplink throughput are 152.37~$\mathrm{Mbps}$ and 24.16~$\mathrm{Mbps}$, respectively, and multi-server RTT is 44~$\mathrm{ms}$. Hughes and Intelsat both use multi-orbit satellite communication systems, delivering median download throughput of 84.55~$\mathrm{Mbps}$ and 61.61~$\mathrm{Mbps}$, respectively.  Viasat, which uses GEO satellites, provided a downlink throughput of 50.38~$\mathrm{Mbps}$. Intelsat delivered an uplink throughput of 9.96~$\mathrm{Mbps}$, while Hughes and Viasat's uplink throughput remains around 1~$\mathrm{Mbps}$.

%It is worth reminding that t
The communication distance is the key contributor to RTT. At the highest elevation angles, a LEO satellite at 550 $\mathrm{km}$ altitude has nearly 60 times lower RTT than a GEO satellite at 35,786 $\mathrm{km}$.
Therefore, Starlink exhibits a RTT of 44 $\mathrm{ms}$, whereas all other providers report significantly higher RTT ranging from 703 $\mathrm{ms}$ to 757 $\mathrm{ms}$, primarily due to the use of multi-orbit architectures, MEO and GEO satellite systems.
 
\if{0}
At the time of writing, the aircraft equipped with the Starlink aero terminal have collectively completed over 200,000 flights, accumulated 540,000 flight hours, and traveled a total distance of 270 million miles. However, to the best of our knowledge, we are not aware of any scientific publication that evaluates the network characteristics of Starlink in-flight connectivity.  There is a clear need to
conduct a thorough in-flight measurement campaign. Our work addresses this shortcoming with intracontinental and intercontinental in-flight measurements.

\fi
%There is a clear need to conduct a thorough in-flight measurement campaign.
At the time of writing, the aircraft equipped with the Starlink aero terminal have collectively completed over 200,000 flights and traveled a total distance of 270 million miles. However, to the best of our knowledge, we are not aware of any scientific publication that evaluates the Starlink in-flight connectivity performance.  Our work addresses this shortcoming with intracontinental and intercontinental measurements.
\section{Experimental Setup and Dataset}
\label{sec:sec3}
We conducted measurements on two separate flights, during different days and airlines across three continents. This section discusses the experimental location, setup and provides an overview of the collected measurement dataset.

\subsection{Locations and Setup}
The first measurement  (denoted as Flight I) was carried out on May 18, 2025, during a HA449 flight (Hawaiian Airlines using Airbus A330) from Honolulu, U.S., to Izumisano, Japan, over the Pacific Ocean. The second measurement  (denoted as Flight II) took place on July 11, 2025, during LH2466 flight (Operated by Air Baltic using Airbus A220-300 on behalf of Lufthansa) from Munich, Germany to Helsinki, Finland, flying over the Baltic Sea. Airbus A330 and Airbus A220-300 can carry up to 278 and 148 passengers, respectively. During these measurements, both flights were nearly fully occupied. The exact number of passengers who have used Starlink's in-flight internet is not publicly available. However, according to a Viasat survey of 11,000 respondents, 79\%  of passengers connect to in-flight Internet when it is available\footnote{Viasat  (https://www.viasat.com/perspectives/aviation/2023/how-todays-passengers-expect-seamless-connectivity/)}. %We are not aware of the exact number of passengers who used Starlink in-flight Internet. According to the Viasat passenger experience survey of 11,000 people, 79\% of passengers connected to in-flight Wi-Fi when available.

\begin{figure}[!t]
\centering
\includegraphics*[width=0.5\textwidth]{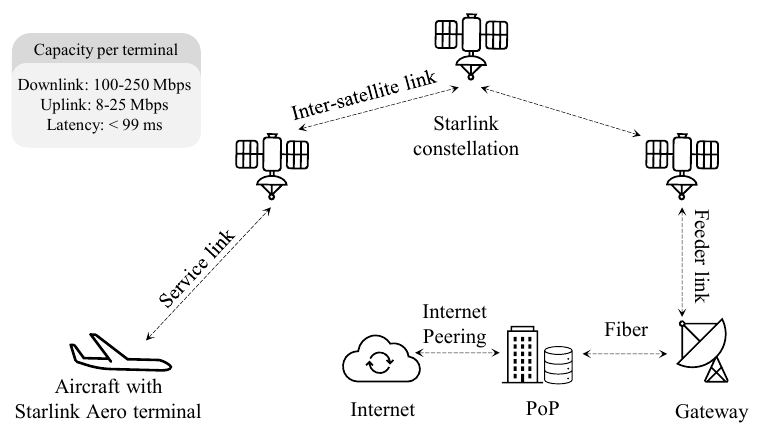}
\caption{Network illustration of Starlink in-flight Internet connectivity.}
\label{fig_2}

\vspace{-12pt}
\end{figure}

Fig. \ref{fig_2} shows the network architecture of Starlink in-flight connectivity. Our measurements were performed on commercial flights without introducing any modification to the in-flight Internet setup. According to Starlink, each terminal comes with two Wi-Fi access points for providing coverage inside the aircraft.  In open ocean and polar regions with absent or sparse ground stations, the Starlink uses optical space lasers for ISLs to connect aircraft with ground stations. During Flight I, the aircraft's aero terminal was connected to a dedicated Starlink ground station in Tokyo, Japan. During Flight II, we pinged dns.google (8.8.8.8), and routing path analysis revealed that Starlink was using a ground station in Frankfurt, Germany, to connect with the Point of Presence (PoP) and the Internet. Considering this, in our results, we are only reporting the RTT of the satellite link from a passenger device to the ground station, excluding the terrestrial networking.
%100.64.0.1 

% \textcolor{red}{Asad: can you please explain this figure a bit further? Let's explain how data travels from the user in the aircraft to a dedicated server using a ground station and PoP, etc. We can also say that our ping results only account for the latency between the aircraft and the ground station. }

\subsubsection{Software} We used three different tools to collect the network characteristics data. First, we used \verb|Ping| to measure the RTT between the client (a passenger device on the aircraft) and a dedicated Starlink ground station. As in \cite{Asad_Starlink}, the term RTT is used as the two-way latency between a client and a dedicated server. In this paper, we use the terms ping, RTT and latency interchangeably. Second,  \verb|tracert| is used to identify the routing path between the client and the ground station. To make a fair comparison with results in Ookla Speedtest Q1 2025 report \cite{ookla2025starlink}, we leveraged the widely known \verb|Ookla Speedtest| application to measure downlink and uplink throughput as well as RTT. Additionally, we also used \verb|Ookla Speedtest| tool to test 4$\mathrm{K}$ video streaming. %capabilities to anticipate real-world application performance.

\subsubsection{Hardware} We collected measurement data using typical passenger devices, including a MacBook Air for Flight I and a smart smartphone (iPhone 15 Pro Max) and an HP EliteBook laptop during tests in Flight II. These devices used in-flight Wi-Fi to connect with the onboard router, which was further connected to the Starlink aero terminal installed on the aircraft's rooftop. Particularly, a smartphone was used to run the \verb|Ookla Speedtest| application, while laptops measured the \verb|Ping|. Typically, Starlink routers use Wi-Fi 5 and Wi-Fi 6 with dual-band (3 $\times$ 3 MIMO) and tri-band (4 $\times$ 4 MU-MIMO) capabilities, respectively. However, the technical details of the aircraft on Wi-Fi routers have not been publicly disclosed by Starlink. We assume that the impact of the onboard Wi-Fi system on overall performance is minimal compared to the satellite link, which is a typical bottleneck. Regarding the static Starlink setup, we used a flat high-performance terminal located in Pisa, Italy, and \verb|Ookla Speedtest| to obtain the network characteristics dataset.

\begin{table}[!t]
\caption{An overview of the flights and measured Starlink datasets.\label{tab:tabII}}
\centering
  \resizebox{\columnwidth}{!}{\begin{tabular}{|l|c|c|c|c|c|}
\hline
\textbf{Test} & \textbf{Location} &\textbf{Continent}& \textbf{Date} &\textbf{Ping} & \textbf{Ookla}\\
\textbf{scenario} &&  &  &\textbf{}& \textbf{Speedtest}\\
\hline
\multirow{2}{*}{Flight I} & Pacific &Asia, &\multirow{2}{*}{18.05.2025}&\multirow{2}{*}{19513}&  \multirow{2}{*}{-}\\

& Ocean &North America&\multirow{2}{*}{}&\multirow{2}{*}{}&  \multirow{2}{*}{}\\
\hline
\multirow{2}{*}{Flight II} & Baltic Sea, &\multirow{2}{*}{Europe}&\multirow{2}{*}{11.07.2025}&\multirow{2}{*}{1610}&  \multirow{2}{*}{86}\\

&  Finland&&\multirow{2}{*}{}&\multirow{2}{*}{}&  \multirow{2}{*}{}\\
\hline
Ground &Pisa & Europe& 27.07.2025&-& 60\\

\hline
\end{tabular}}
\vspace{-10pt}
\end{table}

\subsection{Dataset}
This subsection discusses the flight and Starlink network measured datasets listed in Table~\ref{tab:tabII}.
\subsubsection{Flight data} To observe the impact of aircraft location, speed, and altitude on the Starlink performance, we used the flight track log dataset from the FlightAware\footnote{FlightAware (https://www.flightaware.com/)}. This allows us to analyze network characteristics as a function of aircraft location, altitude, and speed. In Flight~I, the measurements were conducted at an altitude between 35,000 $\mathrm{feet}$ and 40,000~$\mathrm{feet}$, excluding the landing, taxiing, and parking phases.  The measurement took 5.5 hours, and it was conducted between the post-takeoff and pre-landing in-flight period when it was allowed to use a laptop onboard. As a result, dataset is approximately centered at the midpoint of the eight and a half hours flight. Regarding Flight II, the altitude varies, and each measurement sample is fully aligned with aircraft coordinates, altitude, and speed. Notably, FlightAware does not report the flight data for altitudes below 100 $\mathrm{feet}$, which includes the landing, taxiing, and parking stages. To complement the FlightAware dataset for Flight II, we extrapolated aircraft altitude and speed during landing, taxiing, and parking phases. 
\subsubsection{Starlink data} We measured network characteristics for the three scenarios as mentioned in Table~\ref{tab:tabII}. The Flight~I dataset comprises 19,513 \verb|Ping| samples corresponding to 5.5 hours of RTT measurements over the Pacific Ocean. This dataset enables the analysis of RTT behavior during an intercontinental flight as the aircraft travels away from Honolulu, U.S., and approaches the ground station located in Tokyo, Japan. The \verb|Ping| results from Flight II  contain 1,610 samples corresponding to the RTT between the aircraft and the ground station in Frankfurt, Germany. During the same flight, we collected a total of 86 \verb|Ookla Speedtest| samples over the Baltic Sea and Finland.  To make a fair comparison with the Starlink residential terminal, we obtained a total of 60 \verb|Ookla Speedtest| samples from the terminal in Pisa.

%\section{Measurement Dataset}
%\label{sec:sec4}
\section{Measurement Results}
\label{sec:sec4}
This section discusses the flight data, \verb|Ookla Speedtest|, and \verb|Ping| results. It delves deeper into the various factors that drive RTT. Additionally, the section also includes the video streaming quality test results.% in terms of maximum resolution, load time, and percentage of buffering.
\begin{figure}[!t]
\centering
\includegraphics*[width=0.5\textwidth]{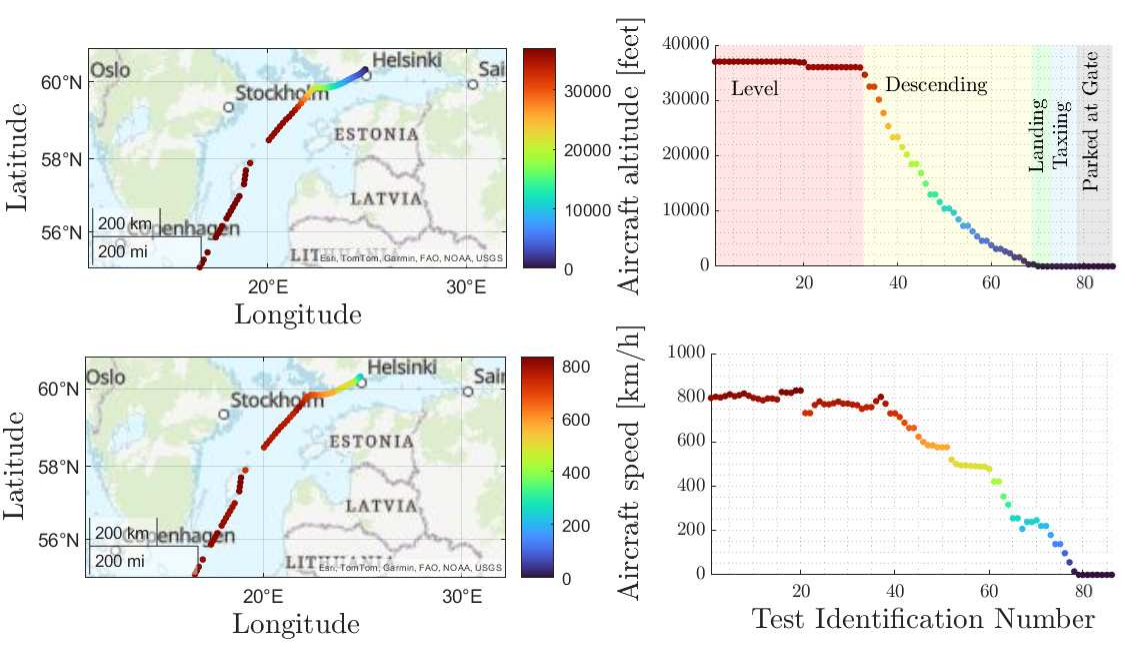}
\caption{Aircraft location, altitude, and speed during Flight II.  This data corresponds to the time of Ookla Speedtest results reported in Fig. \ref{fig_4}.}
\label{fig_3}
\end{figure}
\subsection{Flight Data Overview}
\begin{figure}[!t]
\centering
\includegraphics*[width=0.5\textwidth]{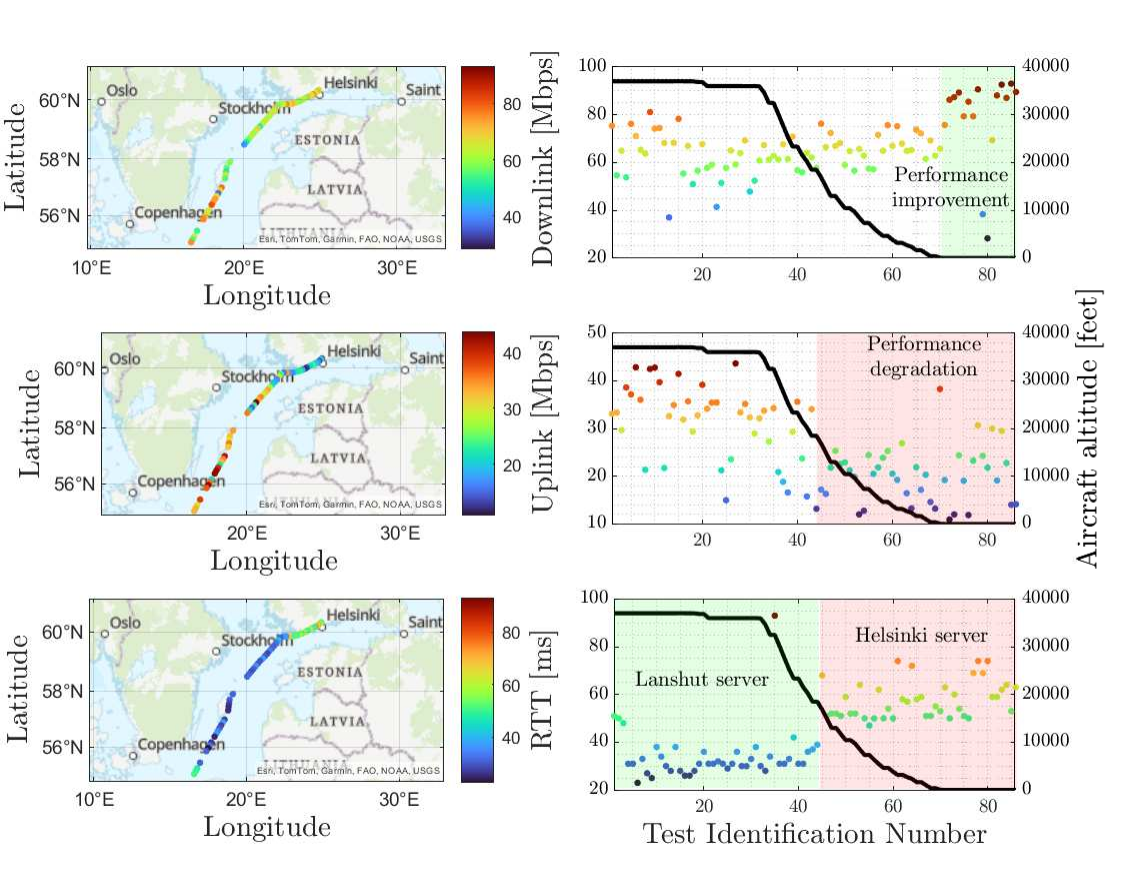}
%\caption{Ookla Speedtest results for Flight II over the Baltic Sea and while coasting into Finland.}
\caption{Ookla Speedtest results for downlink and uplink throughput and RTT during Flight II.}
\label{fig_4}

\vspace{-15pt}
\end{figure}

It is worth reminding that for the Flight I, the altitude remained between 35,000 $\mathrm{feet}$ and 40,000 $\mathrm{feet}$. Regarding Flight II, Fig. \ref{fig_3} depicts the aircraft altitude and speed in the air corridor above the Baltic Sea and while coasting into Finland for landing at Helsinki airport. These aircraft altitude and speed results correspond to the exact time of \verb|Ookla Speedtest| results reported in Fig. \ref{fig_4}. During the first 32 measurements labeled as test identification number~(TIN),  the aircraft altitude and speed were above 36000~$\mathrm{feet}$ and 770~$\mathrm{km/h}$, respectively. After TIN 32, both altitude and speed dropped when the aircraft started descending for landing at Helsinki airport. Since maintaining high speed is a fundamental requirement for a sustained flight, the decrease in altitude occurred more abruptly than the speed reduction. At TIN 45, the altitude and speed reached below 17,000~$\mathrm{feet}$ and 625~$\mathrm{km/h}$, respectively. The aircraft landed during TIN 70, and the remaining 16 measurements were conducted while taxiing on the runway and parked at the gate.

\subsection{Ookla Speedtest}
Fig. \ref{fig_4} illustrates the \verb|Ookla Speedtest| results for downlink, uplink, and RTT. The downlink throughput trend remains the same for the majority of the measurements, with the fluctuations around 65~$\mathrm{Mbps}$. There was no noticeable impact of aircraft speed or altitude on the downlink. However, downlink throughput improves after landing during TIN 70 to 86, when the majority of passengers were preparing to leave the aircraft. Particularly, passengers started to leave the aircraft around TIN 80. We presume that when fewer passenger devices were active, the network may have allocated more downlink resources to our device. Alternatively, it is also possible that many passengers began using cellular connections, e.g., 4G/5G, after landing, thereby decreasing the load on Starlink.

On the contrary, uplink throughput declines with the aircraft altitude and speed. This behavior is visible after TIN 45.  We anticipate that this could be due to the following reasons: (i) there were clouds at lower altitudes that might have caused signal attenuation;  (ii) possibility of more uplink traffic from the passenger onboard for making calls when closer to the airport and (iii) the network control unit (NCU) may have used less Effective Isotropic Radiated Power (EIRP) at lower altitudes. Particularly, aircraft NCU is required to use a low EIRP at lower altitudes to avoid interfering with the terrestrial networks. For example, in the 900~$\mathrm{MHz}$ band, an aircraft uses -10.5~$\mathrm{dBm/200~kHz}$ at 8,000~$\mathrm{meter}$ altitude, which drops to -19~$\mathrm{dBm/200~kHz}$ at 3,000~$\mathrm{meter}$ altitude \cite{micallef2009lband}. We assume that Starlink uses a similar approach.

During the aircraft descent, there was a sudden jump in RTT that occurred when the aircraft's altitude dropped below 17,000~$\mathrm{feet}$. After TIN~44, this sharp increase is mainly due to a change in server location.  \verb|Ookla Speedtest| automatically changed the server location from Landshut, Germany to Helsinki, Finland. Our \verb|tracert| result reveals that the aircraft was connected to a ground station in Germany. During TIN~45 to 86, \verb|Ookla Speedtest| used a server in Helsinki, the communication path between the ground station in Germany and \verb|Ookla Speedtest| server in Helsinki introduced the additional propagation delays that resulted in higher RTT.

%several factors contribute to this difference as follows: (i) The aircraft throughput was equally distributed among the active users; (ii)

\begin{figure}[!t]
\centering
\includegraphics*[width=0.5\textwidth]{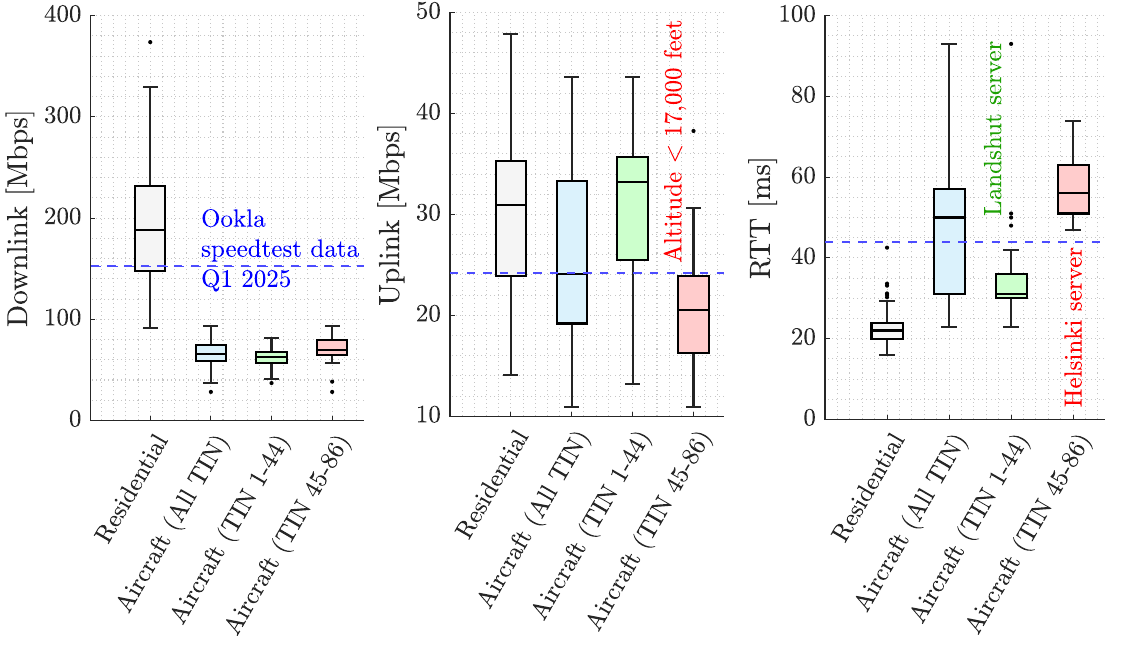}
\caption{Comparison of Starlink ground terminal and in-flight performance obtained by Ookla Speedtest.}
\label{fig_5}

\vspace{-15pt}
\end{figure}

Fig. \ref{fig_5} compares the median performance of \verb|Ookla Speedtest| for a Starlink residential terminal at the University of Pisa, Italy, and the Flight II. For the sake of comparison, the blue dashed lines illustrate the Starlink median in-flight performance as reported in Ookla Speedtest data for Q1 2025 \cite{ookla2025starlink}. A single user device on an aircraft experiences a median downlink of 65~$\mathrm{Mbps}$ while a residential terminal delivers 188~$\mathrm{Mbps}$. Ookla Speedtest data for Q1 2025 shows an in-flight median downlink of 152.4~$\mathrm{Mbps}$. We hypothesize that the aircraft's onboard router may have limited the maximum throughput per user device to distribute the aircraft's total throughput among passengers in a fair manner. The median uplink throughput of all TINs is 24~$\mathrm{Mbps}$, which is the same as  Ookla Speedtest data for Q1 2025 and 6~$\mathrm{Mbps}$ lower than that of the residential terminal. Notably, the median uplink throughput of TIN 1 to 44 when aircraft altitude was above  17,000~$\mathrm{feet}$ is 33~$\mathrm{Mbps}$, which is 3~$\mathrm{Mbps}$ better than the residential terminal. On the contrary, TINs 44-86, which were conducted below 17,000~$\mathrm{feet}$, have a 20~$\mathrm{Mbps}$ median uplink throughput. In all cases, the in-flight median RTT is higher than that of Ookla Speedtest data for Q1 2025 and terrestrial terminal RTT. Particularly, all the TINs have a median RTT of 50~$\mathrm{ms}$ while residential terminals exhibit only 22~$\mathrm{ms}$.  This is mainly due to a sudden increase in RTT when \verb|Ookla Speedtest| automatically changed the server from Landshut, Germany, to Helsinki, Finland.  The TIN 1 to 44 corresponds to the Landshut server with a median RTT of  31~$\mathrm{ms}$. The remaining measurements have a median RTT of 56~$\mathrm{ms}$ while \verb|Ookla Speedtest| connected to a server in Helsinki.

\subsection{Tracert}
During Flight II, we deployed \verb|Ping| using the dns.google (8.8.8.8) server. The \verb|tracert| tool allowed us to examine how a packet travels from a user device on an aircraft to this server. The first hop shows the IP address of the aircraft Wi-Fi router, and after 18 $\mathrm{ms}$, the packet enters the terrestrial network at the second hop, which is the Starlink ground station in Frankfurt, Germany.  On average, a packet takes 9 $\mathrm{ms}$ in the terrestrial part of the network between the ground station and the dns.google server.  We excluded the RTT of the terrestrial part of the network from the results report in Fig.~\ref{fig_6} and Fig.~\ref{fig_7}.%This latency is due to the terrestrial part of the network, which we excluded from the results report in Fig. \ref{fig_6} and Fig. \ref{fig_7}.

\if{0}
\begin{tcolorbox}[colback=white, colframe=black, boxrule=0.5pt, arc=2pt, left=1mm, right=1mm, top=1mm, bottom=1mm, width=\columnwidth]
{\footnotesize
\begin{alltt}
Tracing route to dns.google [8.8.8.8]
over a maximum of 30 hops:

1   2 ms    1 ms    1 ms  10.2.0.1
2   25 ms   18 ms   23 ms  100.64.0.1\footnote{Starlink ground station in Frankfurt, Germany}
3   57 ms   73 ms   66 ms  172.16.251.154
4   22 ms   28 ms   28 ms  [206.224.65.208]\footnote{undefined.hostname.localhost}
5   27 ms   22 ms   22 ms  [206.224.70.85]\footnote{undefined.hostname.localhost}
6   22 ms   25 ms   30 ms  149.19.109.49
7   20 ms   21 ms   26 ms  142.251.48.235
8   19 ms   36 ms   35 ms  172.253.64.119
9   47 ms   26 ms   22 ms  dns.google [8.8.8.8]

Trace complete.
\end{alltt}
}
\end{tcolorbox}
\fi
\subsection{Ping}
%Fig. \ref{fig_6} shows \verb|Ping| results as a function of aircraft location and TIN. 

\begin{figure}[!t]
\centering
\includegraphics*[width=0.5\textwidth]{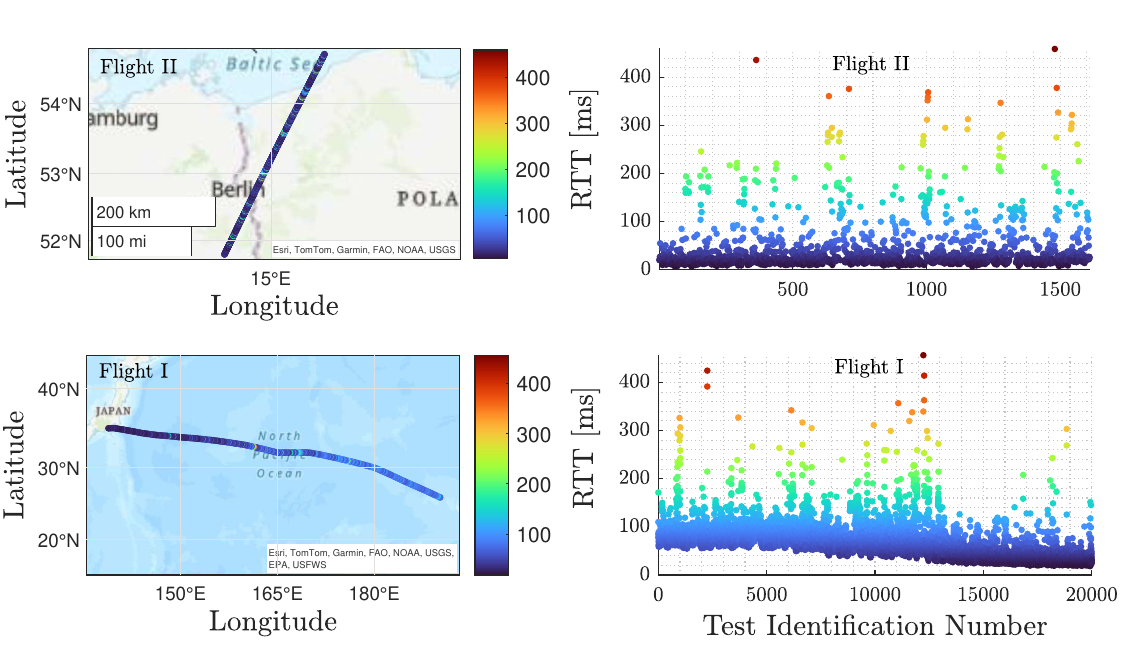}
\caption{RTT results obtained by Ping for Flight I and Flight II.}
\label{fig_6}

\vspace{-15pt}
\end{figure}

\begin{figure}[!t]
\centering
\includegraphics*[width=0.5\textwidth]{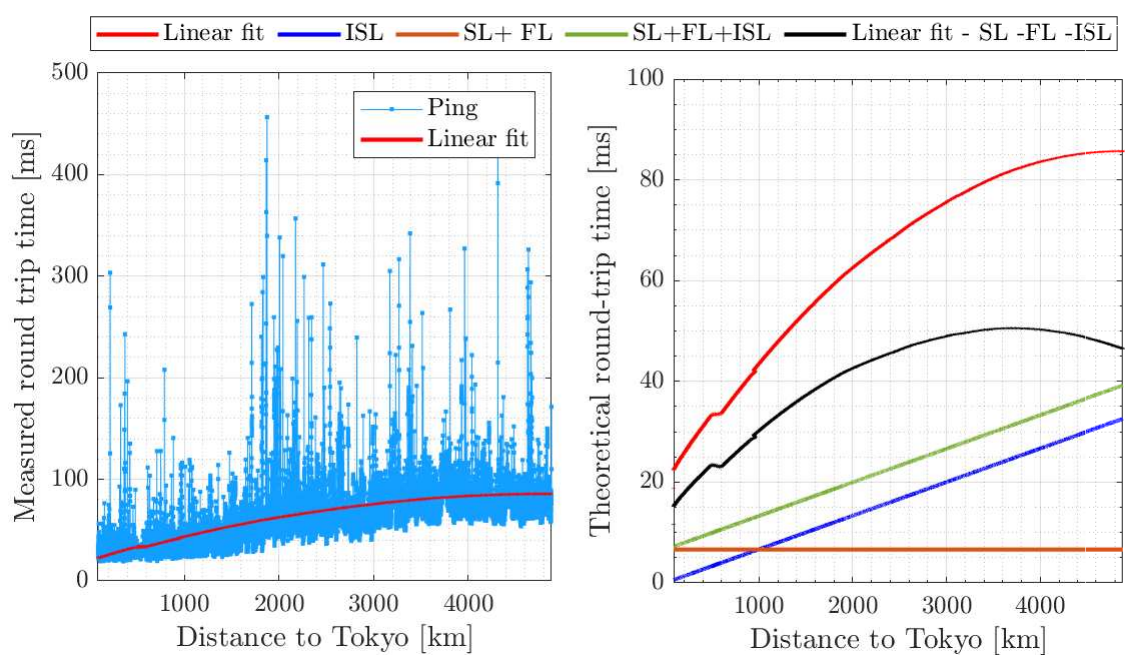}
\caption{The left-hand side of the figure shows the measured RTT by Ping between the aircraft and the Starlink ground station in Tokyo. The right-hand side demonstrates the different key factors that drive theoretical RTT.}
\label{fig_9}

\vspace{-20pt}
\end{figure}

The left-hand side of Fig. \ref{fig_6} illustrates RTT results obtained using \verb|Ping| as a function of aircraft coordinates. The upper plot shows Flight II \verb|Ping| measurements corresponding to 30 minutes, equivalent to around 1610 RTT samples marked by TID as demonstrated on the right-hand side of Fig. \ref{fig_6}. The lower plot presents 19,513  RTT samples obtained by \verb|Ping| for 5.5 hours over the Pacific Ocean during Flight I. One can see that RTT decreases when the aircraft approaches Japan. Particularly, the last TINs have significantly lower RTT than the starting TINs.  We dive deeper into it in Fig. \ref{fig_9}.

%There is no major trend in latency measured by \verb|Ping|.
\begin{figure}[!t]
\centering
\includegraphics*[width=0.5\textwidth]{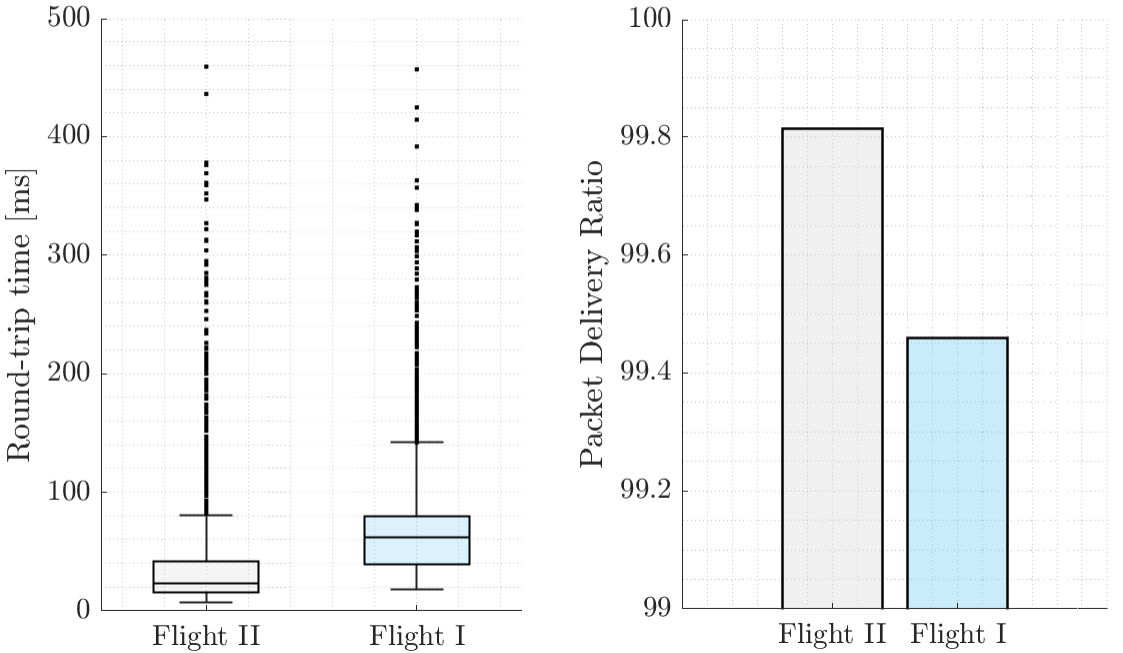}
\caption{ Comparison of RTT and PDR obtained by Ping for both flights. }
\label{fig_7}

\vspace{-15pt}
\end{figure}
The left-hand side of Fig. \ref{fig_7} compares the \verb|Ping| measurements from two different flights on different continents. Flight~I and Flight~II experienced median RTT of 62 $\mathrm{ms}$ and 23 $\mathrm{ms}$, respectively. During Flight I,  we \verb|Ping| a Starlink ground station in Tokyo. The median distance between aircraft and Tokyo was around 2,450 $\mathrm{km}$, which contributed an additional two-way propagation delay of 16.3 $\mathrm{ms}$, resulting in an overall higher RTT as evident in the left-hand side of Fig. \ref{fig_7}. 
On the contrary,  the aircraft's aero terminal was connected to a Starlink ground station in Frankfurt, Germany, during Flight~II. The median distance between aircraft and Frankfurt was around 562 $\mathrm{km}$. Therefore, the Flight II median RTT is significantly lower than that of Flight I.
The right-hand side of Fig. \ref{fig_7} shows the packet delivery ratio (PDR) which remains above 99.4\% for both flights.

As illustrated in Fig. \ref{fig_9}, the communication distance significantly contributes to the RTT. Particularly, the left-hand side of the figure depicts the Flight~I RTT measurements collected by \verb|Ping| as a function of aircraft distance to Tokyo, where we were pinging to the Starlink ground station. One can see that the measured RTT increases with the distance to the ground station. The right-hand side of Fig. \ref{fig_9} dives deeper into different factors that drive the theoretical RTT in the Starlink network.  We used Starlink two-line element data and the MATLAB satellite communications toolbox to calculate the median RTT of the service link (SL), which accounts for two-way propagation delay between the user and the satellite, and the feeder link (FL), accounting for two-way propagation delay from satellite to ground station. Both SL and FL have median cumulative RTT of 6.62 $\mathrm{ms}$.  We assume that the ISL RTT is affected by the two-way propagation delay between the first satellite nearest the aircraft to the last satellite closest to the ground station. ISLs introduce a maximum additional 32 $\mathrm{ms}$ to theoretical RTT.

% We calculate this difference as $\mathrm{Linear fit - SL - FS -  ISL}$.
One can see that the theoretical RTT due to the cumulative sum of $\mathrm{SL}$, $\mathrm{FS}$, and $\mathrm{ISL}$ is lower than the linear fit of the measured RTT.  Particularly, the linear fit of measured RTT was 19 $\mathrm{ms}$ closest to Tokyo, and it increased to 86~$\mathrm{ms}$ when located at a distance of 4900 $\mathrm{km}$.  However, the actual difference between the linear fit of measured RTT and theoretical RTT as $\mathrm{Linear\ fit - SL - FS -  ISL}$ is 47 $\mathrm{ms}$ at the maximum distances. We assume several factors, such as queueing at satellite nodes or delayed scheduling and routing, are contributing to this difference.  According to Starlink, the use of ISL can introduce additional delay, resulting in higher RTT. This may occur due to FL congestion or network routing decisions.  

\begin{table}[!t]
\caption{4K video testing by Ookla Speedtest application.\label{tab:tabIII}}
\centering
  \resizebox{\columnwidth}{!}{\begin{tabular}{|l|c|c|c|c|c|}
\hline
\textbf{Test} &\textbf{Time} &\textbf{Maximum} &\textbf{4K}& \textbf{Load} &\textbf{Buffering}\\
\textbf{Number} &\textbf{GMT}&\textbf{resolution}& \textbf{capabilities} &\textbf{}  &\textbf{percentage}\\
\hline
Test I &8:15 PM&2160p & Yes& 497 ms&0\%\\
Test II &8:15 PM&2160p & Yes& 437 ms&0\%\\
Test III &8:17 PM&2160p & Yes& 335 ms&0\%\\
Test IV &8:19 PM&2160p & Yes& 491 ms&0\%\\

\hline
\end{tabular}}

\vspace{-15pt}
\end{table}

\subsection{Video testing}
\verb|Ookla Speedtest| application measures the performance of video streaming in terms of maximum resolution, load time, and percentage of buffering. Table \ref{tab:tabIII} shows the results for the quality of the video stream. In all tests, network quality supported maximum resolution of 2160p, also known as 4K streaming, buffering was 0\%, and load time ranged between 335 $\mathrm{ms}$ to 497 $\mathrm{ms}$.

\if{0}
\begin{figure}[!t]
\centering
\includegraphics*[width=0.5\textwidth]{4Kstreamesults.pdf}
\caption{Video testing by Ookla Speedtest application.}
\label{fig_10}
\end{figure}

\fi
%\balance
\section{Conclusion}
\label{sec:sec5}
The commercial satellite communication solutions, including Starlink, are poised to become essential for delivering reliable, high-speed internet access in airborne environments. This study explores the significant impact that LEO satellites are expected to have on connectivity within the aviation sector. Our measurements obtained using the Starlink in-flight service during two commercial flights across three continents provide interesting observations and highlight the trade-offs.  In future work, we plan to expand the in-flight measurement campaign to collect more comprehensive data using advanced network characteristics monitoring tools. Particularly, we are interested in the Arctic region, where the Starlink constellation is sparse and weather conditions are harsh during the winter. 

\if{0}
\section*{Acknowledgments}
This research was supported by Business Finland through the  6G-SatMTC projects and partially supported by the Italian Ministry of Education and Research (MUR) in the framework of the FoReLab project (Departments of Excellence) and European Union -Next Generation EU under the Italian National Recovery and Resilience Plan (NRRP), Mission 4, Component 2, Investment 1.3, CUPE83C22004640001, partnership on "Telecommunications of the Future" (PE00000001 -program "RESTART").

\fi
%\textcolor{red}{TO DO}
\bibliographystyle{IEEEtran}
\bibliography{Reference}

\begin{thebibliography}{10}
\providecommand{\url}[1]{#1}
\csname url@samestyle\endcsname
\providecommand{\newblock}{\relax}
\providecommand{\bibinfo}[2]{#2}
\providecommand{\BIBentrySTDinterwordspacing}{\spaceskip=0pt\relax}
\providecommand{\BIBentryALTinterwordstretchfactor}{4}
\providecommand{\BIBentryALTinterwordspacing}{\spaceskip=\fontdimen2\font plus
\BIBentryALTinterwordstretchfactor\fontdimen3\font minus \fontdimen4\font\relax}
\providecommand{\BIBforeignlanguage}[2]{{%
\expandafter\ifx\csname l@#1\endcsname\relax
\typeout{** WARNING: IEEEtran.bst: No hyphenation pattern has been}%
\typeout{** loaded for the language `#1'. Using the pattern for}%
\typeout{** the default language instead.}%
\else
\language=\csname l@#1\endcsname
\fi
#2}}
\providecommand{\BIBdecl}{\relax}
\BIBdecl

\bibitem{10.1145/3614204.3616108}
J.~Garcia \emph{et~al.}, ``Multi-timescale evaluation of starlink throughput,'' in \emph{Proceedings of the 1st ACM Workshop on LEO Networking and Communication}, ser. LEO-NET '23.\hskip 1em plus 0.5em minus 0.4em\relax New York, NY, USA: Association for Computing Machinery, 2023, p. 31–36.

\bibitem{10.1145/3589334.3645328}
N.~Mohan \emph{et~al.}, ``A multifaceted look at starlink performance,'' in \emph{Proc. of the ACM Web Conf. 2024}, ser. WWW '24.\hskip 1em plus 0.5em minus 0.4em\relax New York, NY, USA: Association for Computing Machinery, 2024, p. 2723–2734.

\bibitem{Asad_Starlink}
\BIBentryALTinterwordspacing
M.~Asad~Ullah \emph{et~al.}, ``Starlink in northern {Europe}: A new look at stationary and in-motion performance,'' 2025. [Online]. Available: \url{https://arxiv.org/abs/2502.15552}
\BIBentrySTDinterwordspacing

\bibitem{Dominic_2025}
D.~Laniewski, E.~Lanfer, and N.~Aschenbruck, ``Measuring mobile starlink performance: A comprehensive look,'' \emph{IEEE Open Journal of the Communications Society}, vol.~6, pp. 1266--1283, 2025.

\bibitem{Humphreys}
T.~E. Humphreys \emph{et~al.}, ``Signal structure of the starlink ku-band downlink,'' \emph{IEEE Transactions on Aerospace and Electronic Systems}, vol.~59, no.~5, pp. 6016--6030, 2023.

\bibitem{Asad_Starlink_Weather}
\BIBentryALTinterwordspacing
M.~Asad~Ullah \emph{et~al.}, ``Impact of weather on satellite communication: Evaluating starlink resilience,'' 2025. [Online]. Available: \url{https://arxiv.org/abs/2505.04772}
\BIBentrySTDinterwordspacing

\bibitem{mafakheri2023edge}
B.~Mafakheri \emph{et~al.}, ``Edge intelligence in 5g and beyond aeronautical network with leo satellite backhaul,'' in \emph{2023 Joint Eur. Conf. Netw. Commun. \& 6G Summit}.\hskip 1em plus 0.5em minus 0.4em\relax IEEE, 2023, pp. 579--584.

\bibitem{ookla2025starlink}
\BIBentryALTinterwordspacing
Ookla. (2025, Jun.) Starlink elevates in-flight wi-fi performance. Accessed: 2025-07-25. [Online]. Available: \url{https://www.ookla.com/articles/starlink-elevates-in-flight-wi-fi}
\BIBentrySTDinterwordspacing

\bibitem{qatar2025starlink}
\BIBentryALTinterwordspacing
{Qatar Airways}. (2025, Jul.) Qatar airways leads industry with completion of starlink installation programme for boeing 777s, delivering wi-fi speeds of up to 500 mbps per aircraft. Accessed: 2025-07-25. [Online]. Available: \url{https://www.qatarairways.com/press-releases}
\BIBentrySTDinterwordspacing

\bibitem{Parada}
\BIBentryALTinterwordspacing
R.~Parada \emph{et~al.}, ``Enabling continuous 5g connectivity in aircraft through low earth orbit satellites,'' 2025. [Online]. Available: \url{https://arxiv.org/abs/2504.07262}
\BIBentrySTDinterwordspacing

\bibitem{micallef2009lband}
\BIBentryALTinterwordspacing
J.~Micallef \emph{et~al.}, ``{L-band interference scenarios characterisation},'' Eurocontrol, Technical Report P1031D005, Aug. 2009. [Online]. Available: \url{https://www.eurocontrol.int/sites/default/files/2019-05/25082009-lcis-s1tos5-compatibility-scenarios-v10.pdf}
\BIBentrySTDinterwordspacing

\end{thebibliography}
\newpage
\end{document}